# The Analytical General Solution of the Riccati Equation and Its Intrinsic Functions

Ji-Xiang Zhao
College of Information Engineering, China Jiliang University, Hangzhou, China
zhaojixiang@cjlu.edu.cn, ORCID: 0000-0003-2426-8741

**Abstract:** A sufficient and necessary condition for the integrability of the Riccati equation is established using the f elementary quadrature method. Based on this condition, an analytical general solution involving free variables (parameters) is presented, along with two intrinsic functions and their entanglement relations. This results can be extended to second-order linear ordinary differential equations and provide possible mathematical criteria for modern physics.

## 1. Introduction

The Riccati equation

$$\frac{dy}{dx} = f_2(x)y^2 + f_1(x)y + f_0(x),\ y = y(x) \tag{1}$$

is the simplest nonlinear ordinary differential equation and only one that can be converted into linear equation. Both in the classical theory of differential equations and in the related branches of modern nonlinear science, there is an urgent need to obtain the solution [1-3]. But more than 300 years have passed since Count Jacopo Francesco Riccati discovered the equation, except in some special cases, the problem remains an open one [4, 5]. It is the purpose of this paper to obtain analytic general solution of the Riccati equation.

## 2. Preliminaries

We begin by introducing an existing result.

**Lemma** [5] If the coefficients of Riccati equation

$$y' = a(x)y^2 + b(x)y + c(x),\quad a(x)c(x) \neq 0 \tag{2}$$

satisfy conditions: (1) $c = c_1 + c_2$; (2) $b^2 - 4ac_1 > 0$ and

$$(3)\quad c_2 = \frac{d}{dx}\left(\frac{-b \pm \sqrt{b^2 - 4ac_1}}{2a}\right)$$

then the general solutions of equation (2) can be exactly given as

$$y_\pm(x) = \frac{e^{\pm\int\sqrt{b^2-4ac_1}dx}}{c_0 - \int ae^{\pm\int\sqrt{b^2-4ac_1}dx}dx} + \frac{-b \pm \sqrt{b^2 - 4ac_1}}{2a} \tag{3}$$

where $c_0$ is an integration constant(the proof see appendix).

From this, we obtain the following necessary and sufficient condition for integrability:

**Theorem.1** There exit differential functions $u(x)$, $v(x)$, $\phi(x)$ and $\varphi(x)$, such that

$$\frac{1}{u}\left(f_0 + f_1 v + f_2 v^2 - \varphi v^2 - v'\right) = \left(R\frac{v}{u}\right)' \tag{4a}$$

and

$$2f_2 v + f_1 - \frac{u'}{u} = \phi v \tag{4b}$$

then the Riccati equation (1) is integrable and general solution can be given by

$$\begin{aligned} y(x) &= \frac{ue^{\pm\int v\sqrt{\phi^2-4f_2\varphi}dx}}{c_0 - \int f_2 ue^{\pm\int v\sqrt{\phi^2-4f_2\varphi}dx}dx} + (1+R)v \\ &= \frac{e^{\int Idx}}{c_0 - \int f_2 e^{\int Idx}dx} + u\int\frac{1}{u}\left[f_0 + (\phi - f_2 - \varphi)v^2\right]dx \end{aligned} \tag{5}$$

where $R(x) \triangleq \dfrac{-\phi \pm \sqrt{\phi^2 - 4f_2\varphi}}{2f_2}$ and $I(x) = f_1 + 2f_2 u\displaystyle\int\frac{1}{u}\left[f_0 + (\phi - f_2 - \varphi)v^2\right]dx$.

**Proof**: (**Sufficiency**) The substitution $y(x) = u(x)z(x) + v(x)$ brings equation (1) into

$$z' = f_2 uz^2 + \left(2f_2 v + f_1 - \frac{u'}{u}\right)z + \frac{1}{u}\left(f_0 + f_1 v + f_2 v^2 - v'\right) \tag{6}$$

we introduce differentiable functions $\phi(x)$ and $\varphi(x)$, such that equation (4b) holds,

then equation (6) can be rewritten as follow

$$z' = f_2uz^2 + \phi vz + \frac{1}{u}\left(f_0 + f_1v + f_2v^2 - v' - \varphi v^2 + \varphi v^2\right) \tag{7}$$

set $a = f_2u$ , $b = \phi v$ , $c_1 = \frac{\varphi v^2}{u}$ and $c_2 = \frac{1}{u}\left(f_0 + f_1v + f_2v^2 - \varphi v^2 - v'\right)$, by the **Lemma**, if equation (4a) holds, we have

$$z(x) = \frac{e^{\pm\int v\sqrt{\phi^2 - 4f_2\varphi}dx}}{c_0 - \int f_2ue^{\pm\int v\sqrt{\phi^2 - 4f_2\varphi}dx}dx} + R\frac{v}{u} \tag{8}$$

note that using the identity $\frac{v'}{u} = \left(\frac{v}{u}\right)' + \frac{u'}{u^2}v$ , equation (4a) can be written as follows

$$\left[(1+R)\frac{v}{u}\right]' = \frac{1}{u}\left[f_0 + \left(f_1 - \frac{u'}{u}\right)v + (f_2 - \varphi)v^2\right]$$

$$\overset{f_1 - \frac{u'}{u} = \phi v - 2f_2v}{=} \frac{1}{u}\left[f_0 + (\phi - f_2 - \varphi)v^2\right]$$

integrating the above equation gives

$$(1+R)v = u\int\frac{1}{u}\left[f_0 + (\phi - f_2 - \varphi)v^2\right]dx \tag{10}$$

furthermore, with the help of equation (4b), we obtain

$$\pm v\sqrt{\phi^2 - 4f_2\varphi} = f_1 - \frac{u'}{u} + 2f_2u\int\frac{1}{u}\left[f_0 + (\phi - f_2 - \varphi)v^2\right]dx \tag{11}$$

thus, by $y(x) = uz + v$ , we get to formula (5).

(**Necessity**) Differentiating formula (5) with respect to $x$, we obtain

$$\begin{aligned} y' = f_0 + f_1&\left[\frac{e^{\int I(x)dx}}{c_0 - \int f_2e^{\int I(x)dx}dx} + u\int\frac{1}{u}\left[f_0 + (\phi - f_2 - \varphi)v^2\right]dx\right] \\ &+ f_2\left[\frac{e^{\int I(x)dx}}{c_0 - \int f_2e^{\int I(x)dx}dx}\right]^2 + \frac{2f_2e^{\int I(x)dx}}{c_0 - \int f_2e^{\int I(x)dx}dx}\left[u\int\frac{1}{u}\left[f_0 + (\phi - f_2 - \varphi)v^2\right]dx\right] \\ &+ (f_2 - \varphi)v^2 - \left(f_1 - \frac{u'}{u}\right)\left[-v + u\int\frac{1}{u}\left[f_0 + (\phi - f_2 - \varphi)v^2\right]dx\right] \end{aligned} \tag{12}$$

if

$$
\begin{aligned}
&(f_2-\varphi)v^2-\left(f_1-\frac{u'}{u}\right)\left[-v+u\int\frac{1}{u}\left[f_0+(\phi-f_2-\varphi)v^2\right]dx\right] \\
&=f_2\left[u\int\frac{1}{u}\left[f_0+(\phi-f_2-\varphi)v^2\right]dx\right]^2
\end{aligned}
\tag{13}
$$

holds, equation (12) becomes

$$
\begin{aligned}
y'=&f_2\left[\frac{e^{\int I(x)dx}}{c_0-\int f_2e^{\int I(x)dx}dx}+u\int\frac{1}{u}\left[f_0+(\phi-f_2-\varphi)v^2\right]dx\right]^2 \\
&+f_1\left[\frac{e^{\int I(x)dx}}{c_0-\int f_2e^{\int I(x)dx}dx}+u\int\frac{1}{u}\left[f_0+(\phi-f_2-\varphi)v^2\right]dx\right]+f_0
\end{aligned}
$$

this shows that formula (5) is the general solution of Riccati equation (1) under the constraint equation (13). For equation (13), with the help of equation (4b), it can be rewritten as a generalized quadratic algebraic equation

$$
\begin{aligned}
&f_2\left[u\int\frac{1}{u}\left[f_0+(\phi-f_2-\varphi)v^2\right]dx\right]^2 \\
&+(\phi-2f_2)v\left[u\int\frac{1}{u}\left[f_0+(\phi-f_2-\varphi)v^2\right]dx\right]-(\phi-f_2-\varphi)v^2=0
\end{aligned}
$$

there are two roots

$$
u\int\frac{1}{u}\left[f_0+(\phi-f_2-\varphi)v^2\right]dx=\frac{(2f_2-\phi)v\pm v\sqrt{\phi^2-4f_2\varphi}}{2f_2}
$$

dividing both sides by $u$, and differentiating, notice that $\left(\frac{v}{u}\right)'=\frac{v'}{u}-v\frac{u'}{u^2}$, by some simple operations, the above equation can be transformed into equation (4a).

**3. General solution**

The analytical general solution of the Ricatti equation is systematically derived using the following theorem.

**Theorem.2** The Riccati equation (1) is integrable by elementary quadrature method and general solution including free variable and parameters is obtained.

**Proof**: (**Sufficiency**) We first solve for $\phi(x)$, $\varphi(x)$, $u(x)$ and $v(x)$ that satisfy equations (4), for which, with the help of equation (4b), equation (4a) can be written

as

$$\left[(1+R)v\right]' = \frac{R(2f_2-\phi)}{(1+R)^2}\left[(1+R)v\right]^2 + f_1\left[(1+R)v\right] + (f_2-\varphi)v^2 + f_0 \tag{14}$$

We assume that $(1+R)v = \dfrac{\alpha}{\dfrac{R(2f_2-\phi)}{(1+R)^2}}$ (where $\alpha$ is a nonzero constant ), yields

$$v(x) = \frac{\alpha(1+R)}{R(2f_2-\phi)} \tag{15}$$

and equation (14) is simplified to

$$\left[\frac{R(2f_2-\phi)}{(1+R)^2}\right]' + (\alpha+f_1)\left[\frac{R(2f_2-\phi)}{(1+R)^2}\right] + \frac{1}{\alpha}\left[(f_2-\varphi)v^2+f_0\right]\left[\frac{R(2f_2-\phi)}{(1+R)^2}\right]^2 = 0 \tag{16}$$

set

$$(f_2-\varphi)v^2 = \lambda \tag{17}$$

where $\lambda(x)$ is an arbitrary function, and from equation (16), we get

$$\frac{R(2f_2-\phi)}{(1+R)^2} = \frac{e^{-\int(\alpha+f_1)dx}}{\beta + \dfrac{1}{\alpha}\displaystyle\int(\lambda+f_0)e^{-\int(\alpha+f_1)dx}dx} \triangleq A(x,\lambda,\alpha,\beta) \tag{18}$$

where $\beta$ is an integration constant. Thus $(1+R)v = \dfrac{\alpha}{A}$, with the help of equation (4b), i.e. $(2f_2-\phi)v = \dfrac{u'}{u} - f_1$, we have

$$\pm v\sqrt{\phi^2-4f_2\varphi} = \frac{2\alpha f_2}{A} + f_1 - \frac{u'}{u} \tag{19}$$

Finally, by substituting $(1+R)v = \dfrac{\alpha}{A}$ and equation (19) into equation (5), we obtain

$$y(x,\lambda,\alpha,\beta) = \frac{e^{\int\left(\frac{2\alpha f_2}{A}+f_1\right)dx}}{c_0 - \displaystyle\int f_2 e^{\int\left(\frac{2\alpha f_2}{A}+f_1\right)dx}dx} + \frac{\alpha}{A} \tag{20}$$

**(Necessity)** We take the derivative of equation (20) and, with the help of equations (15) and (17), obtain

$$y' = f_0 + f_1\left[\frac{e^{\int\left(\frac{2\alpha f_2}{A}+f_1\right)dx}}{c_0 - \int f_2 e^{\int\left(\frac{2\alpha f_2}{A}+f_1\right)dx}dx} + \frac{\alpha}{A}\right] + f_2\left[\frac{e^{\int\left(\frac{2\alpha f_2}{A}+f_1\right)dx}}{c_0 - \int f_2 e^{\int\left(\frac{2\alpha f_2}{A}+f_1\right)dx}dx} + \frac{\alpha}{A}\right]^2 + \frac{\alpha^2}{A} - f_2\left(\frac{\alpha}{A}\right)^2 + (f_2 - \varphi)\left[\frac{\alpha(1+R)}{R(2f_2-\phi)}\right]^2 \tag{21}$$

in fact, with the help of equation (18)

$$\frac{\alpha^2}{A} - f_2\left(\frac{\alpha}{A}\right)^2 + (f_2-\varphi)\left[\frac{\alpha(1+R)}{R(2f_2-\phi)}\right]^2 \overset{\frac{\alpha(1+R)}{R(2f_2-\phi)}=\frac{\alpha}{A(1+R)}}{=} \frac{\alpha^2}{A} - f_2\left(\frac{\alpha}{A}\right)^2 + (f_2-\varphi)\left[\frac{\alpha}{A(1+R)}\right]^2$$

$$= \left(\frac{\alpha}{A}\right)^2\left[A - f_2 + \frac{f_2-\varphi}{(1+R)^2}\right] \overset{A=\frac{R(2f_2-\phi)}{(1+R)^2}}{=} -\left(\frac{\alpha}{A}\right)^2\frac{1}{(1+R)^2}\left(f_2R^2 + \phi R + \varphi\right) = 0$$

and equation (21) is reduced to

$$y' = f_0 + f_1\left[\frac{e^{\int\left(\frac{2\alpha f_2}{A}+f_1\right)dx}}{c_0 - \int f_2 e^{\int\left(\frac{2\alpha f_2}{A}+f_1\right)dx}dx} + \frac{\alpha}{A}\right] + f_2\left[\frac{e^{\int\left(\frac{2\alpha f_2}{A}+f_1\right)dx}}{c_0 - \int f_2 e^{\int\left(\frac{2\alpha f_2}{A}+f_1\right)dx}dx} + \frac{\alpha}{A}\right]^2$$

this shows that formula (20) is the general solution of the Ricatti equation (1).

## 4. Intrinsic functions

It is worth noting that intrinsic functions $\phi\left[\alpha,\beta,\lambda(x)\right]$ and $\varphi\left[\alpha,\beta,\lambda(x)\right]$ can be obtained from equations (17) and (18) and given by

$$\phi\left[\alpha,\beta,\lambda(x)\right] = 2(f_2 - A) \pm \frac{A^2\lambda + \alpha^2(f_2-A)}{\alpha\sqrt{\lambda(f_2-A)}} \tag{22}$$

$$\varphi\left[\alpha,\beta,\lambda(x)\right] = A\left(1 - \frac{A\lambda}{\alpha^2} \pm \frac{2\sqrt{\lambda(f_2-A)}}{\alpha}\right) \tag{23}$$

their entanglement relation is given by equations (16) or (18) that can be rewritten as

$$\frac{R(2f_2-\phi)}{(1+R)^2} = \frac{e^{-\int(\alpha+f_1)dx}}{\beta + \frac{1}{\alpha}\int\left[(f_2-\varphi)\left(\frac{\alpha(1+R)}{R(2f_2-\phi)}\right)^2 + f_0\right]e^{-\int(\alpha+f_1)dx}dx} \tag{24}$$

## 5. Concluding remark

The general solution involving free variable $\lambda(x)$ and parameters $\alpha$, $\beta$ to the Riccati equation is presented in this paper, as well as the intrinsic functions ($\phi, \varphi$) and entangled relation are discovered and given by equations (22-24). It is worth to say, that these results can also be applied to solve the second-order linear homogeneous ordinary differential equation $y''(x)+p(x)y'(x)+q(x)y(x)=0$, which can be transformed to a Riccati equation $z'=-z^2-pz-q$ by the transformation $z(x)=\dfrac{y'(x)}{y(x)}$.

**Appendix**: For general form of the Riccati equation

$$y'=a(x)y^2+b(x)y+c(x),\quad ac\neq 0$$

if the coefficients satisfy: (1) $c=c_1+c_2$ and (2) $c_2=\left(\dfrac{-b\pm\sqrt{b^2-4ac_1}}{2a}\right)'$, then the general solutions can be exactly given as

$$y_{\pm}(x)=\frac{e^{\pm\int\sqrt{b^2-4ac_1}dx}}{c_0-\int ae^{\pm\int\sqrt{b^2-4ac_1}dx}dx}+\frac{-b\pm\sqrt{b^2-4ac_1}}{2a}$$

where $c_0$ is an integration constant.

**Proof**: We rewrite $y'=ay^2+by+c$ as follow

$$\begin{aligned} y'&=ay^2+by+c \\ &=a\left(y+\frac{b+\sqrt{b^2-4ac_1}}{2a}\right)\left(y+\frac{b-\sqrt{b^2-4ac_1}}{2a}\right)+\left(\frac{-b\pm\sqrt{b^2-4ac_1}}{2a}\right)' \end{aligned}$$

further, we can transform it into a *Bernoulli*-type equation

$$\left(y-\frac{-b\pm\sqrt{b^2-4ac_1}}{2a}\right)'\mp\sqrt{b^2-4ac_1}\left(y-\frac{-b\pm\sqrt{b^2-4ac_1}}{2a}\right)=a\left(y-\frac{-b\pm\sqrt{b^2-4ac_1}}{2a}\right)^2$$

and this solution is easy to get.

## Declaration of Interest Statement

The author declares that there is no conflict of interests regarding the publication of this paper.

## Data availability

No data were used for the research described in this paper.